\documentclass[prb,amsmath,showpacs,preprint,floatfix]{revtex4}
\usepackage{graphicx}

\makeatletter
\let\latex@xfloat=\@xfloat
\def\@xfloat #1[#2]{%
  \latex@xfloat #1[#2]%
  \def\baselinestretch{1}%
  \small
}
\makeatother

\begin{document}

%% ---------- TITLE ----------
\title{Electron-electron interaction effects on the photophysics of metallic
single-walled carbon nanotubes}

\author{Zhendong Wang}
\affiliation{Department of Physics, University of Arizona
Tucson, AZ 85721}
\author{Demetra Psiachos}
\affiliation{Department of Physics, University of Arizona
Tucson, AZ 85721}
\author{Roberto F. Badilla}
\affiliation{Department of Physics, University of Arizona
Tucson, AZ 85721}
\author{Sumit Mazumdar}
\affiliation{Department of Physics, University of Arizona
Tucson, AZ 85721}

\date{\today}

%
% 73.22.-f   Electronic structure of nanoscale materials: clusters,
% nanoparticles, nanotubes, and nanocrystals
%
% 78.67.Ch   Optical properties of Nanotubes
%
% 71.35.-y   Excitons and related phenomena

%%% ---------------------------------- 
%%  ---------        abstract --------
%%-------------------------------------

\begin{abstract}

%%SM10/26 - adding a sentence
Single-walled carbon nanotubes are strongly correlated systems with large Coulomb repulsion between
two electrons occupying the same $p_z$ orbital. 
Within a molecular Hamiltonian appropriate for correlated $\pi$-electron systems,
we show that optical excitations polarized parallel to the nanotube axes in the so-called
metallic single-walled carbon nanotubes 
are to excitons. Our calculated absolute exciton
energies in twelve different metallic single-walled carbon nanotubes, with diameters in the
range 0.8 - 1.4 nm, are in nearly quantitative agreement with experimental results. We have
also calculated the absorption spectrum for the (21,21) single-walled
carbon nanotube in the E$_{22}$ region. Our calculated spectrum gives an excellent fit to the
experimental absorption spectrum. In all cases our calculated exciton  
binding energies are only slightly smaller than 
those of semiconducting nanotubes with comparable diameters, in contradiction to results obtained
within the {\it ab initio} approach, which predicts much smaller binding energies.
We ascribe this difference to the difficulty
of determining the behavior of systems with strong on-site Coulomb interactions within theories
based on the density functional approach.
As in the semiconducting nanotubes we predict in the metallic nanotubes 
a two-photon exciton above the lowest longitudinally polarized exciton that can be detected by 
ultrafast pump-probe spectroscopy. We also predict a subgap absorption polarized perpendicular to the nanotube axes   
below the lowest longitudinal exciton, blueshifted from the exact midgap by
electron-electron interactions. 
\end{abstract}


\maketitle

\section{Introduction}

Correlated electron systems often exhibit behavior that are substantively different from what
is expected within one-electron (1-e) theory. In particular, the classification of materials as simple metals
or semiconductors breaks down for sufficiently strong electron-electron (e-e) interactions. 
The effects of e-e interactions are particularly strong in low dimension, and
carbon-based quasi-one-dimensional (quasi-1D) systems such as $\pi$-conjugated polymers, 
semiconducting and conducting charge-transfer solids, and carbon nanotubes 
commonly exhibit
novel behavior ascribed to e-e interactions. Although it is by now generally accepted that Coulomb interactions
between the $\pi$-electrons are strong in single-walled carbon nanotubes (SWCNTs), they continue to be
classified as metallic (M-SWCNTs) and semiconducting (S-SWCNTs), based on the predictions
of 1-e theory. Thus, SWCNTs with chirality indices
(n,m) are commonly referred to as
metallic if (n-m) = 3j, where j is an integer including zero, and semiconducting otherwise.
Schematic $\pi$-electron tight-binding band structures of the armchair (n=m) and nonarmchair
(n$\neq$m, including
m=0) M-SWCNTs are shown in Fig.~1. The innermost valence and conduction bands (VB and CB, respectively)
have linear dispersions and meet at
Dirac points, which constitute the Fermi points here. The crossing innermost bands are missing
in the S-SWCNTs; otherwise their bandstructures are similar to that in Fig~1(b). 
We continue to use the nomenclature based
on 1-e theory for simplicity in what follows, with the recognition that simple classifications
of SWCNTs may not be entirely meaningful.
%% ----------------------------
%%          Fig. 1
%% ----------------------------
\begin{figure}
\centering
\includegraphics[width=3.375 in]{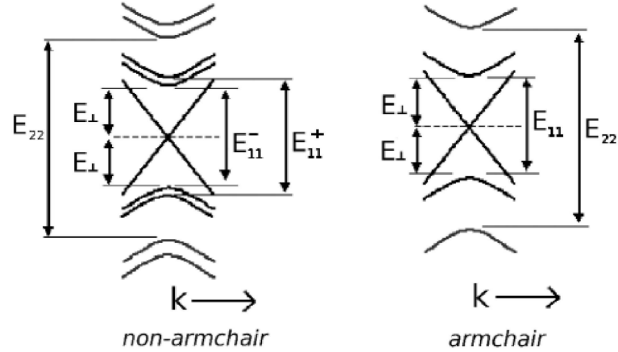}
\caption{(a) Schematics of the tight-binding band structures of M-SWCNTs and
longitudinal and transverse optical transitions within one-electron theory.
Splittings due to the trigonal warping are indicated for nonarmchair NTs. The
same E$_{11}$ and E$_{22}$ transitions occur in the S-SWCNTs, which do not have the
inner crossing bands.
}
\end{figure}

In recent years,
there has been a strong interest in the consequences of e-e interactions on the photophysics
of S-SWCNTs. The bulk of the existing literature is
on optical absorptions polarized parallel to the NT axes, where e-e interactions lead to exciton formation.
The exciton character of the longitudinally polarized absorptions in S-SWCNTs
\cite{Ando97a,Lin00a,Kane03a,Spataru04a,Chang04a,Perebeinos04a,Zhao04a,Wang06a,Dresselhaus07a,Scholes07a} is now
firmly established. Nonlinear absorption\cite{Zhao06a} and two-photon induced fluorescence
\cite{Wang05a,Dukovic05a,Maultzsch05a}
have demonstrated that the binding energy of the lowest longitudinal optical exciton
in S-SWCNTs is substantial relative to the optical gap. Research on optical absorptions
polarized {\it perpendicular} to the NT axes has been less extensive 
\cite{Zhao04a,Miyauchi06a,Uryu06a,Wang07a,Kilina08a}, but the consequences of e-e interactions
here are even more dramatic. Within 1-e theory, the perpendicularly polarized absorption
occurs exactly at the center of the two lowest longitudinally polarized absorptions (hereafter
E$_{11}$ and E$_{22}$, see Fig.~1). The experimentally
observed strong blueshift of the perpendicularly polarized absorption \cite{Miyauchi06a} 
to near E$_{22}$ is due to e-e interactions \cite{Zhao04a,Uryu06a,Wang07a,Kilina08a}.

There exists also a considerable body of theoretical 
\cite{Balents97,Egger97,Kane97,Krotov97,Yoshioka99,Odintsov99,Bunder07} and experimental
\cite{Tans97,Tans98,Bockrath97,Bockrath99} literature on the effects of e-e interactions on the
M-SWCNTs, that until recently had focused mostly on transport behavior.
%Within 1-e theory, an (n,m) SWCNT is metallic if (n-m) is three 
%times an integer. Schematic band structures of the armchair (n=m) and nonarmchair (n$\neq$m, including
%m=0) M-SWCNTs are shown in Fig.~1. The innermost valence and conduction bands (VB and CB, respectively) 
%have linear dispersions characteristic of one-dimensional (1D) systems and meet at
%Dirac points, which constitute the Fermi points here.
Screening of the interactions between the $\pi$-electrons in these 1D systems is weak, and the 
lowest excitations in M-SWCNTs
have been shown to correspond to those of a Luttinger liquid (LL) rather than a Fermi liquid.
Indeed, it has been claimed \cite{Balents97} that the lowest excitations of an (n,n) armchair M-SWCNTs can
be approximately mapped onto those of two-leg ``Hubbard ladders'' \cite{Dagotto96} 
with an effective on-site Hubbard repulsion
$U_{eff} \sim U/n$, where $U$ is the repulsion between two electrons occupying the same 
$p_z$ carbon orbital.
Given that $U$ is substantial in carbon-based systems \cite{Zhao04a,Wang06a}, this would suggest that
the narrowest armchair nanotubes
with diameters $d \simeq 1$ nm are likely Mott-Hubbard semiconductors with both charge- and 
spin-gaps. \cite{Balents97,Krotov97,Yoshioka99} 
%(simple estimates suggest that even the (10,10)
%armchair nanotube with $d=1.38$ nm is a semiconductor with a small but nonzero charge-gap \cite{Balents97}.) 
Although 
excitations in nonarmchair M-SWCNTs are more complex, it is believed that the low energy physics of these
are the same as in the armchair tubes. Finally, while the above discussions concerning the Mott-Hubbard 
semiconductor nature of the narrowest M-SWCNTs focused
on the short-range component of the e-e interactions, the charging energy of a tube is determined
primarily by the long-range component, which has also been shown to be weakly screened \cite{Egger97}.
Fitting the experimental charging energy of a M-SWCNT \cite{Tans97} with a 1/$|x|$ potential, 
for example,
requires a dielectric constant of only 1.4 \cite{Egger97}.

It is in this context that we examine theoretically the photophysics of M-SWCNTs here. 
We are
concerned not about the lowest excitations involving the electrons occupying the innermost bands
in Fig.~1, but optical transitions in the visible region. 
Electronic transitions leading to optical absorptions within 1-e theory are indicated in Fig.~1. 
In addition to the VB-to-CB transitions that are
polarized
parallel to the NT axes, we expect also midgap transitions polarized perpendicular to the NT axes, based 
on our experience with the S-SWCNTs. \cite{Zhao04a,Wang07a} 
Only the absorptions parallel to the NT-axis have been experimentally investigated in
M-SWCNTs so far. \cite{Strano03a,Telg04a,Fantini04a,Fantini07a,Wang07b}
%%SM10/26 - note modification
In view of the weak screening
of the e-e interactions in M-SWCNTs (see above), we expect the ``large Hubbard $U$'' description to be
appropriate here, {\it even if these systems are conducting and are not Mott-Hubbard semiconductors}
Note that unlike in true 1D, the Hubbard $U$ has to be larger than a critical value before a metal-to-insulator
transition will occur in graphene. Conducting behavior thus is not a signature of reduced $U$.
%% up to here.
Taken together with with the large atomic U scenario, the 1:1 correspondence of the 
VB-to-CB transitions in Fig.~1 to those in the S-SWCNTs then suggests that
photoexcitations in M-SWCNTs are to excitons with binding energies that are perhaps
comparable to those in the
S-SWCNTs. This conjecture is, however, in strong contradiction to existing theoretical results.
\cite{Deslippe07a} Within the latter method the ground state is determined using an {\it ab initio}
approach, which is followed by the determination of the quasiparticle energies
within the GW approximation and the solution of the Bethe-Salpeter
equation of the two-particle Green's function. 
This technique has claimed that binding energies in M-SWCNTs are
an order of magnitude smaller than those in S-SWCNTs with comparable diameters. A recent work has
also claimed that the experimental E$_{22}$ absorption of the (21,21) armchair M-SWCNT 
can be fit well within the 
{\it ab initio}
theory, and that the exciton binding energy in this system is only 0.05 eV \cite{Wang07b}. The
absence of two-photon induced flourescence in M-SWCNTs (because of the inner VB and CB) has prevented the
direct measurement of exciton binding energies. It then becomes imperative to investigate the photophysics
of M-SWCNTs theoretically using other approaches. 

In the present paper, we report the results of 
many-body calculations of the photophysics of M-SWCNTs, based on a molecular Hamiltonian that has 
previously yielded quantitatively accurate results for the absolute exciton energies, exciton binding 
energies and nonlinear absorption in S-SWCNTs \cite{Zhao04a,Zhao06a,Wang06a,Wang07a}.
The exciton binding energies we obtain for M-SWCNTs are considerably larger than those 
found in reference
\onlinecite{Deslippe07a}. In agreement with the earlier LL theories, our results indicate that
screening of the electron-hole interactions in M-SWCNTs is considerably weaker than in
conventional metals.
%We believe that our results are congruent with the earlier LL description of the lowest excitations in
%M-SWCNTs \cite{Balents97,Kane97,Egger97,Krotov97,Yoshioka99,Odintsov99,Bunder07}. These systems are
%not true metals, and the conventional concepts of screening do not apply to them.  

In section II we present our $\pi$-electron Hamiltonian and indicate how the parameters of the
Hamiltonian are obtained. We then give a brief 
justification of the choice of our parameters.
In section III.A we present our theoretical results for linear and nonlinear
absorptions in the M-SWCNTs. Our results for the absolute exciton energies are in excellent
agreement with experiments for all twelve M-SWCNTs that we have studied. Our calculated exciton binding
energies are much larger than those predicted within the {\it ab initio} theory. In section III.B we compare
our calculated absorption spectrum of the (21,21) M-SWCNT with the experimental spectrum. \cite{Wang07b} 
Again, excellent agreement between the
theoretical and experimental absorption spectra is obtained. Finally, in section III.C we present our
predicted theoretical absorption spectra polarized perpendicular to the NT axes. As with the S-SWCNTs,
\cite{Zhao04a,Uryu06a,Wang07a,Kilina08a}
the perpendicularly polarized absorptions show dramatic effects of e-e interactions. Unlike
in the S-SWCNTs, though, the lowest perpendicularly polarized absorptions will occur {\it below}
the lowest logitudinal absorption in the M-SWCNTs. In section IV we present our conclusions, focusing on
the difference between our results and those obtained within the {\it ab initio} approach. \cite{Deslippe07a}
%and on the implication for the electronic structures of M-SWCNTs.

\section{$\pi$-electron model and its parametrization}

We investigate theoretically the photophysics of M-SWCNTs within the same $\pi$-electron
Pariser-Parr-Pople (PPP) \cite{Pariser53a} model that we have used
for the S-SWCNTs \cite{Zhao04a,Wang06a} and planar $\pi$-conjugated polymers \cite{Chandross97a},
\begin{equation}
\label{H_PPP}
\begin{split}
H = &-t\sum_{\langle ij \rangle, \sigma} 
(c_{i,\sigma}^\dagger c_{j,\sigma}+ H.C.) +
U \sum_{i} n_{i,\uparrow} n_{i,\downarrow}\\ 
&+ \sum_{i<j} V_{ij} (n_{i}-1)(n_{j}-1)
\end{split}
\end{equation}
where $c^{\dagger}_{i,\sigma}$ creates a $\pi$-electron of spin $\sigma$ on
carbon atom $i$, $n_{i,\sigma} = 
c^{\dagger}_{i,\sigma}c_{i,\sigma}$ is
the number of electrons on atom $i$ with spin $\sigma$ and
$n_{i} = \sum_{\sigma}n_{i,\sigma}$ is the total number of electrons on atom
$i$. Here $t$
is the nearest neighbor one-electron hopping, 
$U$ and $V_{ij}$ are the
on-site and intersite Coulomb interactions.
We parametrize $V_{ij}$
as \cite{Zhao04a,Wang06a,Chandross97a}
\begin{equation}
\label{parameters}
V_{ij}=\frac{U}{\kappa\sqrt{1+0.6117 R_{ij}^2}} 
\end{equation}
where $R_{ij}$ is the distance between carbon atoms $i$ and $j$ in
\AA, and $\kappa$ is the background dielectric
constant. Since full many-body calculations are not possible within
Eq.~\ref{H_PPP}, we use the single-configuration interaction (SCI), which retains
all matrix elements between single-excitations from the Hartree-Fock (HF) ground state.
Calculations reported below are for 60 or more unit cells, with open boundary conditions
\cite{Zhao04a,Wang06a}.

The three independent parameters within Eq.~\ref{H_PPP} are
$t$, $U$ and $\kappa$. The nearest neighbor hopping integral is widely accepted to be 2.4 eV
in planar $\pi$-conjugated systems. \cite{Chandross97a} The hopping in SWCNTs is smaller because
of the curvature, which decreases the overlaps between neighboring p$_z$ orbitals.
A smaller $t$ of 2.0 eV for S-SWCNTs was 
determined from careful fitting of the experimental data. \cite{Wang06a}  
Since the curvature effects in M-SWCNT's are the same
as in S-SWCNTs, we use the same $t=2.0$ eV as in the S-SWCNTs. Not surprisingly,
the Hubbard
on-site repulsion $U$ is found to be the same in both $\pi$-conjugated polymers \cite{Chandross97a} and
S-SWCNTs \cite{Zhao04a,Wang06a,Wang07a}, viz., 8 eV, which would place both these classes of materials
among strongly correlated-electron systems. In the context of a different class of 1D
correlated-electron materials, organic charge-transfer solids, it has been shown by numerous authors 
in the past that the short-range e-e interaction, in particular the Hubbard $U$, remains
practically unchanged between the $\frac{1}{2}$-filled band semiconductors and 
the non-$\frac{1}{2}$-filled conductors 
\cite{Soos74,Torrance79,Hubbard78,Mazumdar83,Mazumdar86}. This conclusion has been substantiated by
more recent work \cite{Claessen02a,Bozi08a} and is also in agreement with theories of high temperature
superconductors, within which the undoped Mott-Hubbard semiconductors and the doped conductors and
superconductors are generally assumed to have the same $U$. Based on prior work, we therefore expect
the Hubbard $U$ to be the same in M-SWCNTs and S-SWCNTs, and
use $U=8$ eV in our calculations repored here.

The long range interaction $V_{ij}$ in M-SWCNTs, however, can be different
from S-SWCNTs due to screening, and this is taken into account
by modifying $\kappa$.
We arrive at
the appropriate $\kappa$ by comparing the experimental lowest longitudinal
exciton energies in three different M-SWCNTs: (8,8) armchair, (12,0) zigzag and
(9,6) chiral with PPP-SCI energies, calculated using multiple values of $\kappa$.
In Table I we show our comparisons of the calculated and experimental quantities for the
(8,8), (12,0) and (9,6) M-SWCNTs.
The two nonarmchair M-SWCNTs, in which E$_{11}$ splits into a lower E$_{11}^-$ and an upper
E$_{11}^+$ due to trigonal warping \cite{Strano03a}, provide rigorous tests
of our theory. As seen in the Table, while the $\kappa$ appropriate for
M-SWCNTs is certainly larger than the value 2 used for S-SWCNTs,\cite{Wang06a} 
$\kappa >3$ yields
exciton energies that are too small. The only exception to this is E$_{11}^-$
for the (12,0) NT. Note, however, that (i) this is the narrowest NT considered
(as has been emphasized in reference \onlinecite{Wang06a}, $\pi$-electron theory becomes 
less quantitative for small $d$),
and (ii) even here the best fit to E$_{11}^+$ is with $\kappa=3$. We have therefore chosen
$\kappa=3$ in what follows.

Justification of the choice of our Hamiltonian and our parameters come from two considerations.
First, previous theoretical works on M-SWCNTs have already
emphasized weak screening of e-e interactions in M-SWCNTs.
\cite{Balents97,Egger97,Kane97,Krotov97,Yoshioka99,Odintsov99,Bunder07} Our determination that
$\kappa$ in M-SWCNTs is only slightly larger than that in S-SWCNTs agrees with the conclusion of
reference \onlinecite{Egger97} that fitting the charging energy in M-SWCNTs requires a
relatively small dielectric constant.
Second, in the case of S-SWCNTs, the PPP-SCI approach
has provided the best agreement with experimental absolute exciton energies and exciton binding energies to date
for nanotubes with $d \geq$ 1 nm.
The {\it maximum} difference between our previously calculated and experimental E$_{11}$ for S-SWCNTs
with diameters in this range is  
0.05 eV, while for slightly narrower tubes with $d$ between 0.75 - 1.0 nm, this difference
is 0.1 eV \cite{Wang06a}. Our calculated exciton binding energies of 0.4 to 0.3 eV 
for S-SWCNTs with
$d \sim 0.8 - 1.0$ nm are within 0.04 eV of the experimental quantities on the average
\cite{Wang06a}. Our
calculated energies of absorptions polarized
perpendicular to the NT axes for four different S-SWCNTs with $d \sim$ 1 nm are
also within 0.1 eV of experimental values. \cite{Wang07a} 
%We justify the use of a static homogenous dielectric constant
%based on the results of reference \onlinecite{Egger97}. 
%Finally, as has been emphasized by Hubbard in the context of conducting charge-transfer solids,
%screening effects should emerge in the {\it solution} to the
%correlated-electron Hamiltonian in Eq.~1 and are ``not to be taken into account separately in setting up
%this Hamiltonian.''

%%-----------------------
%%   Table I
%%-------------------------
\begin{table}
\centering
\caption{Calculated and experimental \cite{Strano03a} exciton energies for three M-SWCNTs
with
$t=2.0$ eV and $U=8$ eV, and several different $\kappa$. }
%The experimental results are from Ref. \onlinecite{Strano03a}).}
\label{tab:fit-k-MSWCNTs}
\begin{ruledtabular}
\begin{tabular}{ccccccc}
&
&
&
\multicolumn{2}{c}{$E_{11}^{-}$ (eV)}&
\multicolumn{2}{c}{$E_{11}^{+}$ (eV)}
\tabularnewline \cline{4-5} \cline{6-7}
($n$,$m$) &
d (nm)&
$\kappa$ &
PPP &
Expt. &
PPP &
Expt. \\
\hline
(8,8) & 1.10 & 2.0 & 2.35 & 2.11& \textemdash & \textemdash
\\   &      &2.2&2.26& &\textemdash & \\
   &      & 3.0&2.07& &\textemdash & \\
   &      & 3.5&2.00& &\textemdash & \\
   &      &4.0 &1.94 & &\textemdash & \\
   \hline
   & & & & & &
   \\
   (12,0) & 0.95 &
   2.0&
   2.57&
   2.16&
   2.71&
   2.47\\
   & & 2.2& 2.48& & 2.63& \\
   & & 3.0& 2.28& & 2.49& \\
   & & 3.5& 2.21& & 2.42& \\
   & & 4.0& 2.15& & 2.37& \\
   \hline
   & & & & & &
   \\
   (9,6) & 1.04 & 2.0& 2.35& 2.15& 2.52& 2.22\\
   & & 2.2& 2.30& & 2.45& \\
   & & 3.0& 2.14& & 2.25& \\
   & & 3.5& 2.08& & 2.17& \\
   & & 4.0& 2.03& & 2.12& \\
\end{tabular}
\end{ruledtabular}
\end{table}

\section{Results}

\subsection{Linear and nonlinear absorptions in M-SWCNTs}

In Table \ref{tab:M-CNT-E11} we present our calculated and experimental \cite{Strano03a,Telg04a,Fantini04a,Fantini07a}
absolute energies of the excitons for twelve different M-SWCNT's with $d>0.8$ nm. 
We compare theoretical results mostly against the experimental results of
reference \onlinecite{Strano03a}, 
which is the only work that
reports both E$_{11}^-$ and E$_{11}^+$ for the nonarmchair M-SWCNTs.
We obtain excellent fits to experiments in all cases. Importantly, our  
calculations reproduce almost quantitatively the small energy differences between
E$_{11}^-$ and E$_{11}^+$. 
Our largest deviations, 0.18 eV for E$_{11}^-$ and 0.12 eV for E$_{11}^+$, are for the
(10,1) NT with smallest $d$. 
%(as discussed in reference \onlinecite{Wang06a}, $\pi$-electron
%theory becomes less quantitative as $d$ decreases).
%The overall standard deviations of the theoretical E$_{11}^-$ and E$_{11}^+$ from the 
%experiments are much smaller:
%0.07 eV and 0.05 eV, respectively. 
Theoretical and experimental results agree particularly well
for the (15,0) and the (13,1) NTs, for which the
experimental quantities reported in the different references are close. 
%%-------------------
%%    Table II
%%--------------------

\begin{table*}
\caption{Calculated and experimental exciton energies in M-SWCNTs, and
the calculated binding energies of the excitons}
\label{tab:M-CNT-E11}
\begin{ruledtabular}
\begin{tabular}{ccccccc}
& & \multicolumn{2}{c}{$E_{11}^{-}$ (eV)}&
\multicolumn{2}{c}{$E_{11}^{+}$ (eV)}&
\multicolumn{1}{c}{$E_{b1}$ (eV)}
\\ \cline{3-4}\cline{5-6} \cline{7-7}
($n,m$) &$d$ (nm) &PPP &Expt. &PPP &Expt.$^{a}$ &PPP \\
\hline
(7,7) &0.96 &2.31&2.34$^{a}$,2.43$^{b}$&\textemdash &\textemdash &0.31  \\
(8,8) &1.10 &2.07 &2.11$^{a}$,2.22$^{b}$&\textemdash &\textemdash &0.28 \\
(9,9) &1.24 &1.88 &1.91$^{a}$,2.03$^{b}$,2.02$^{c}$&\textemdash &\textemdash &0.25 \\
(10,10) &1.38 &1.72&1.75$^{a}$,1.89$^{b,c}$&\textemdash &\textemdash &0.23 \\
(10,1) &0.84 &2.51&2.33$^{a}$,2.28$^{b}$,2.38$^{c}$&2.83&2.71&0.30 \\
(9,3) &0.86 &2.47&2.36$^{a}$,2.35$^{b}$,2.43$^{c}$&2.71&2.61&0.29 \\
(8,5) &0.90&2.42&2.37$^{a}$,2.47$^{b,c}$&2.54&2.47&0.29 \\
(12,0) &0.95 &2.28&2.16$^{a,b}$&2.49&2.47&0.27 \\
(10,4) &0.99 &2.22 &2.17$^{a}$,2.22$^{b}$&2.37 &2.33&0.27 \\
(9,6) &1.04 &2.14&2.15$^{a}$,2.23$^{b}$,2.24$^{c}$&2.25&2.22&0.27 \\
(13,1) &1.07 &2.07&2.01$^{a}$, 2.02$^{b}$, 2.06$^{c}$&2.26&2.24&0.25 \\
(15,0) &1.19 &1.91&1.86$^{a,c}$,1.88$^{b}$&2.03&2.06&0.24 \\
\end{tabular}
\end{ruledtabular}
\footnotetext{$^a$From Ref.~\onlinecite{Strano03a}. $^b$From
Refs.~\onlinecite{Fantini04a} and \onlinecite{Fantini07a}.
$^c$From Ref.~\onlinecite{Telg04a}.}
\end{table*}
%-----------------
% end of table II
%-----------------

Table \ref{tab:M-CNT-E11}  also lists our calculated binding energies E$_{b1}$, which we define as
the energy difference between the lower threshold of the
continuum band and the E$_{11}$ (E$_{11}^-$) exciton in armchair (nonarmchair) M-SWCNTs.
Within the SCI approximation, the Hartree-Fock threshold gives the threshold of the continuum.
\cite{Chandross97a,Zhao04a,Wang06a} The 
E$_{b1}$ in all cases are significantly larger than those obtained within 
\textit{ab initio} theory \cite{Deslippe07a}, and are 70-80\% of the exciton binding
energies in S-SWCNTs with similar diameters.\cite{Wang06a} 
%These values are significantly larger than
%the predictions of earlier \textit{ab initio} theory \cite{Deslippe07a}. 
For M-SWCNTs
with $d \sim 1$ nm, for instance, the \textit{ab initio} work had predicted E$_{b1}$ $\sim 0.05$ eV, while
the PPP values are 0.25 - 0.30 eV. 
%We comment on this difference
%later. 
Our ability to reproduce
the small energy differences between E$_{11}^-$ and E$_{11}^+$ gives us confidence about
our calculated E$_{b1}$.

%%---------------------
%%  Figure 2
%%---------------------
\begin{figure}
\centering
\includegraphics[width=3.2 in]{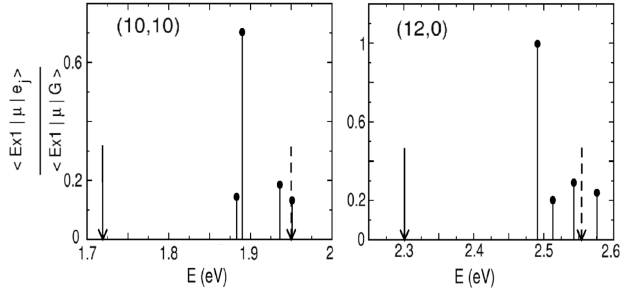}
\caption{Transition dipole couplings between above-gap excited states $j$
and the
optical exciton Ex1, relative to the dipole coupling between Ex1 and the ground state G,
in the (10,10) and (12,0) M-SWCNTs. 
The solid and dashed arrows denote the energy locations
of the optical exciton and the threshold of the continuum band, respectively.
}
\end{figure}

The predicted large E$_{b1}$ can be verified from pump-probe measurments of excited state
absorptions \cite{Zhao06a}.
In Figs.~2(a) and (b) we show the calculated normalized transition dipole couplings between the
lowest optical exciton and higher energy two-photon states in the (10,10) and (12,0)
M-SWCNTs. As in the S-SWCNTs \cite{Zhao06a}, 
there occurs a dominant two-photon exciton that is strongly dipole-coupled to the
optical exciton, and that therefore should be visible as excited state absorption. 
We find similar results in the other metallic NTs. 
The energy difference
between the two-photon exciton and the optical exciton is the lower bound to  E$_{b1}$. 

\subsection{Optical absorption in the (21,21) M-SWCNT}

The absorption spectrum in the E$_{22}$ region of the (21,21) M-SWCNT ($d=2.9$ nm) has recently
been obtained experimentally. \cite{Wang07b}
The absorption band is asymmetric, with weak but significant absorption on the
high energy side of the peak in the absorption (see Fig.~3). Based on comparisons with
the rigidly downshifted symmetric E$_{44}$ absorption spectrum of the (16,15)
S-SWCNT and lineshape analysis, the authors
of this work concluded that E$_{b2}$ in (21,21) M-SWCNT is only 0.05 eV.
As {\it ab initio} calculation for the wide (21,21) M-SWCNT is difficult, the authors used the 
calculated \textit{ab initio} E$_{11}$ transition of the (10,10) S-SWCNT ($d=1.38$ nm) to fit the 
experimental E$_{22}$ absorption of the (21,21) M-SWCNT, since within band theory the two one-electron
gaps have the same origin and are the same in magnitude 
(the absorptions to the exciton and the continuum band were, however, calculated separately
and superimposed in this work). The \textit{ab initio} E$_{b1}$ of the (10,10) M-SWCNT is 
also $\sim$ 0.05 eV,
seemingly supporting the conjecture that the E$_{11}$
exciton of the (10,10) M-SWCNT and the  E$_{22}$ exciton of the (21,21) M-SWCNT
are equivalent even when e-e interactions are significant. Note that our calculated 
E$_{b1}$ in the (10,10) M-SWCNT in 
Table II is, however, significantly larger (0.23 eV), implying that substituting the E$_{11}$
spectrum of the (10,10) NT for the E$_{22}$ spectrum of the (21,21) NT may not be appropriate.
 
We have calculated directly the entire absorption spectrum in the E$_{22}$ region of the
(21,21) M-SWCNT within a single calculation using the PPP-SCI approach. Comparison of the
theoretical and experimental absorption spectra provides a direct test of our theory.
Our calculated E$_{22}$ is 1.75 eV, in good agreement with the experimental
E$_{22}$ of 1.87 eV. \cite{Wang07b} The calculated exciton energy is indeed close to
E$_{11}$ in the
(10,10) M-SWCNT (see Table II).
In Fig.~3 we compare our calculated absorption spectrum, rigidly shifted by
the 0.12 eV energy difference between our calculated and the experimental E$_{22}$,
with the experimental data points of reference \onlinecite{Wang07b}. Apart from this rigid shift,
the fitting is excellent: the calculated spectrum
reproduces both the asymmetric line shape as well as the high energy tail. The latter is not due
to absorption to the continuum band \cite{Deslippe07a}, but is due to weak absorptions to higher
excitons that lie below the continuum band threhold. Similar absorptions to higher excitons are known to contribute to the asymmetric
lineshapes of absorptions within the PPP Hamiltonian, whenever the exciton binding energy
is relatively small \cite{Chandross97a}, and occur also in the
perpendicularly-polarized absorptions in S-SWCNTs
with $d \sim 1$ nm, where the transverse excitons have binding energies of 0.1 - 0.15 eV
(see the experimental absorption spectra in Fig. 3(d) in reference \onlinecite{Miyauchi06a} and
the calculated absorption spectra in Fig.~3 of reference \onlinecite{Wang07a}).
For comparison
to the absorption to an exciton in a S-SWCNT, as was done in reference \onlinecite{Wang07b},
we have superimposed
in Fig.~3 the calculated absorption band in the E$_{22}$ region of the (19,0) S-SWCNT,
again rigidly shifted such that the peaks of the two calculated absorptions match.
According to the prescription of reference \onlinecite{Wang07b}, the threshold
of the E$_{22}$ continuum of the (21,21) M-SWCNT should occur at the energy where the
absorptions of the semiconducting and the metallic NTs begin to diverge, viz., at $\sim$
1.92 eV from Fig.~1(c).
The actual calculated threshold of the continuum, indicated by the arrow in Fig.~3,  
is, however,
at a significantly higher energy. We calculate E$_{b2}$ in
the (21,21) M-SWCNT to be 0.12 eV, nearly half that of the (10,10) M-SWCNT. 

%%SM10/26 - separating to new paragraph
For S-SWCNTs, E$_{b1}$ and E$_{b2}$ for the same
system are comparable. Furthermore, exciton binding energies in S-SWCNTs decrease with
diameter. \cite{Zhao04a,Wang06a} If one assumes both of these
to be true in M-SWCNTs, comparable E$_{b2}$ in the (21,21) M-SWCNT and E$_{b1}$
in the (10,10) M-SWCNT, as calculated within the {\it ab initio} theory are not expected.
The large difference between our calculated
E$_{b1}$ of 0.23 eV in the (10,10) M-SCWNT (see Table II) and E$_{b2}$ of 0.12 eV
in the (21,21) M-SCWNT, 
{\it in spite of the same absolute energies of the corresponding excitons}, in contrast, is 
in agreement with the diameter dependence in the semiconductors. The difference in the two
binding energies is not surprising. The thresholds of the continua in our calculations
correspond to the Hartree-Fock thresholds within Eq.~1. These energies are different for the
(10,10) and (21,21) M-SWCNTs.
even as their tight-binding thresholds are nearly the same.
Although the lowest
excitations in the M-SWCNTs do not necessarily reflect the behavior of the higher energy
excitations, it is interesting that the mapping suggested in reference \onlinecite{Balents97}
predicts a $U_{eff}$ in the (21,21) M-SWCNT that is half the $U_{eff}$ in the (10,10) M-SWCNT.

%% ----------------------------
%%          Fig. 3
%% ----------------------------
\begin{figure}
\centering
\includegraphics[width=3.0 in]{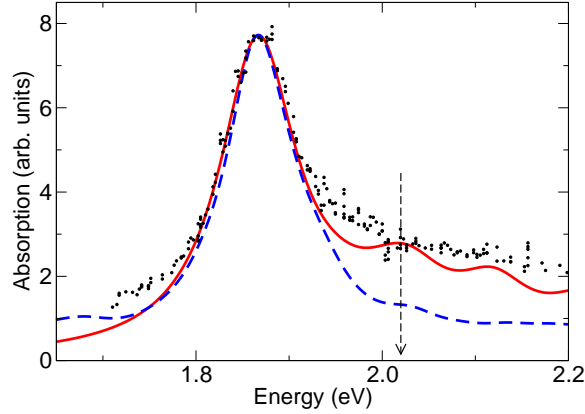}
\caption{
(Color online) Calculated absorption spectrum (red curve) in the E$_{22}$ region of the
(21,21) M-SWCNT, superimposed on the experimental data \cite{Wang07b}
(black dots). The calculated spectrum has been shifted rigidly by 0.12 eV.
The arrow gives the calculated threshold of the continuum band.
The blue dashed curve is the calculated E$_{22}$ absorption of the (19,0) S-SWCNT, shifted rigidly
so that the peaks of the two calculated spectra match.
Linewidths of 0.05 eV and 0.04 eV, respectively, for the (21,21) and (19,0) NT's,
have been used.
}
\end{figure}

\subsection{Perpendicularly polarized absorption in M-SWCNTs}

We now make a verifiable prediction concerning optical absorption polarized 
perpendicular to the NT axes. 
The strong blueshift of the transverse absorption from the exact center of E$_{11}$ and
E$_{22}$ in the S-SWCNTs \cite{Miyauchi06a} is due to e-e interactions
\cite{Zhao04a,Uryu06a,Wang07a,Kilina08a}.   
Degenerate basis functions reached by E$_{12}$ and E$_{21}$
excitations here from new correlated electron eigenstates that are odd and even superpositions
of these basis functions. The redshifted odd superposition is optically forbidden, while the
blueshifted even superposition is optically allowed. \cite{Zhao04a,Wang07a} We anticipate the
degenerate perpendicularly-polarized one-electron transitions in M-SWCNTs
(see Fig.~1) to be also similarly split by e-e interactions,
giving rise to a
redshifted forbidden transition and a blueshifted allowed absorption. The novel feature here, however,
is that the lowest perpendicularly polarized absorption is ``subgap'', occurring below the
lowest longitudinal optical absorption.

In Figs. 4(a) and (b) we have shown our calculated
perpendicularly polarized absorptions for the (7,7) and the (12,0) M-SWCNTs, 
where we have also included the
longitudinal E$_{11}$ absorptions. The subgap perpendicularly polarized absorptions
are blueshifted substantially from the exact midgap. In spite of this strong
Coulomb effect, we find the binding energy of the perpendicular absorption in the 
M-SWCNTs to be nearly zero. 

%%---------------------
%%  Fig 4
%%---------------------
\begin{figure}[htb]
\centering
\includegraphics[width=3.2 in]{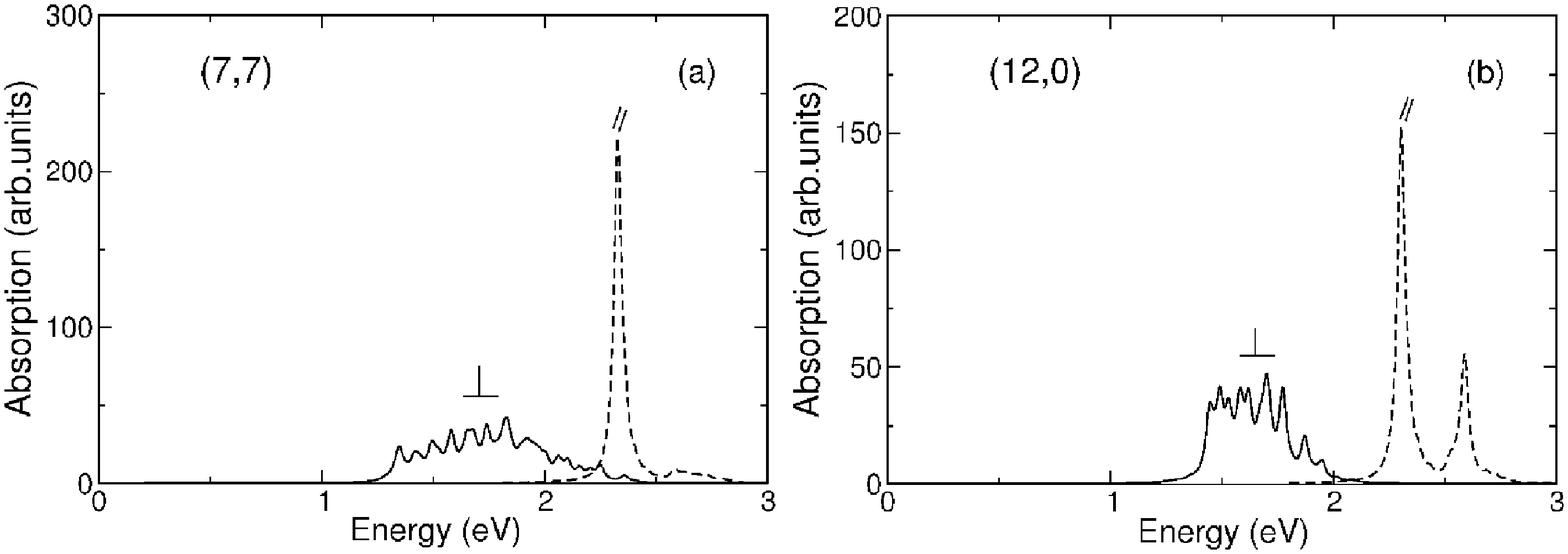}
\caption{Calculated optical absorptions polarized perpendicular to the NT axes in the
(a) (7,7) and (b) (12,0) M-SWCNTs. The E$_{11}$ absorptions are included for
comparison (the splitting of E$_{11}$ in (b) is due to trigonal warping). The
%%SM8/6 - adding
zero frequency Drude absorptions are not shown.
}
\end{figure}

\section {Conclusions}

To conclude, M-SWCNTs are expected to exhibit optical behavior very similar to S-SWCNTs,
with only slightly smaller exciton binding energies. We emphasize that within the PPP
%%SM10/26 - modifying sentence slightly
Hamiltonian of Eq.~1, determining the absolute energy of the exciton and its binding energy are {\it not} 
different problems.
In the limit of large $U$ with only nearest neighbor intersite e-e interaction $V_1$, for example, the exciton 
in a purely 1D system occurs at energy
$U-V_1$ while the conduction band is centered at $U$ \cite{Guo93}. {\it Thus, in this limit, 
once the $U$ is fixed, it is
not possible to obtain the precise exciton energy but incorrect exciton binding energy.}
For moderate $U$, where the hopping term plays a stronger role, it is necessary to also fix the $t$;
but once again, for fixed $U$ and $t$, correct determination of the absolute exciton energy within
Eq.~1 necessarily implies that the continuum band threshold has also been correctly evluated. Based
on our argument in section I that the atomic $U$ is the same in the S-SWCNTs and the M-SWCNTs then,
the excellent fits to the absolute exciton energies in Table II, as well as to the optical absorption
spectrum in Fig.~3, imply that our estimates of the exciton binding energies are correct.
The large E$_b$ implies weak screening of Coulomb interactions. As we have pointed out, weak screening of
e-e interactions in these 1D materials 
\cite{Balents97,Egger97,Kane97,Krotov97,Yoshioka99,Odintsov99,Bunder07} suggests that simple
concepts of metallic screening do not apply.

The discrepancy between
the predictions of the molecular model used here and the {\it ab initio} approach is not unexpected.
Note that even for the S-SWCNTs, the calculated exciton binding energies within the two methods
are widely different, with the {\it ab initio} approach predicting binding energies \cite{Spataru04a} 
that are often twice the experimental values. \cite{Dukovic05a} 
%In contrast, the molecular model reproduces
%the experimental binding energies of S-SWCNTs quantitatively. \cite{Wang06a}
Although it has been suggested that the experimental binding energies reflect
screening of e-e interactions due to intertube interactions, and the true single tube binding energies
are much larger and close to the {\it ab initio} predictions, 
an alternate possibility is that the molecular model, which reproduces experimental
longitudinal {\it and} transverse exciton energies and exciton binding energies quantitatively, is simply
better calibrated to handle systems with large Hubbard interaction. The difficulty of treating
strong on-site e-e interaction within density functional based theories, for instance, is well known. 
\cite{Kotliar06a,Savrasov01a,Cohen08a} 
%Note that within Eq.~1, the
%absolute energy of the exciton and its binding energy are not unrelated. In the limit of
%large U with only nearest neighbor intersite e-e interaction $V_1$, for example, the exciton 
%in 1D occurs at energy
%$U-V_1$ while the conduction band is centered at $U$ \cite{Guo93}. The quantitative fittings of the exciton
%energies in Table II, as well as the fitting of the absorption spectrum of the (21,21)
%M-SWCNT in Fig.~3, both suggest therefore that our estimates of binding energies are correct.

SWCNTs are currently of strong interest because of their potential 
technological applications. Our demonstration that that M-SWCNTs will exhibit photophysics
similar to the semiconductors, even as their transport behavior correspond to that of
unconventional conductors, may introduce new and exciting possibilities.

\section{acknowledgments}
We are grateful to Professors F. Wang and T. F. Heinz for sending us the experimental
absorption data for the (21,21) NT, and to Professors A. Shukla and Z. V. Vardeny for
critical reading of the manuscript. This work was supported by NSF grant number
DMR-0705163.

\end{document}